\documentclass[twocolumn,showpacs,superscriptaddress,amssymb,10pt,pra,floatfix]{revtex4}

\usepackage{graphicx}
\usepackage{dcolumn}
\usepackage{bm}
\usepackage{epsfig}
\usepackage{color}
\usepackage{longtable}

\begin{document}

\preprint{}
%
%
%
%
\title{Elastic scattering of twisted electrons by diatomic molecules}
%
%
%
%

\author{A.~V.~Maiorova}
\affiliation{Center for Advanced Studies, Peter the Great St. Petersburg Polytechnic University, 195251 St. Petersburg, Russia}

\author{S.~Fritzsche}
\affiliation{Helmholtz--Institut Jena, D--07743 Jena, Germany}
\affiliation{Theoretisch--Physikalisches Institut, Friedrich--Schiller--Universit\"at Jena, D--07743 Jena, Germany}

\author{R.~A.~M\"uller}
\affiliation{Physikalisch--Technische Bundesanstalt, D--38116 Braunschweig, Germany}
\affiliation{Technische Universit\"at Braunschweig, D--38106 Braunschweig, Germany}

\author{A.~Surzhykov}
\affiliation{Physikalisch--Technische Bundesanstalt, D--38116 Braunschweig, Germany}
\affiliation{Technische Universit\"at Braunschweig, D--38106 Braunschweig, Germany}

\date{\today \\[0.3cm]}

%
%
%
%

\begin{abstract}
The elastic scattering of twisted electrons by diatomic molecules is studied within the framework of the non--relativistic first Born approximation. In this process, the coherent interaction of incident electrons with two molecular centers may cause interference patterns in the angular distributions of outgoing particles. We investigate how this Young--type interference is influenced by the complex internal structure of twisted beams. In particular, we show that the corkscrew--like phase front and the inhomogeneous intensity profile of the incident beam can strongly modify the angular distribution of electrons, scattered off a single well--localized molecule. For the collision with a macroscopic target, composed of randomly distributed but aligned molecules, the angular--differential cross section may reveal valuable information about
the transverse and longitudinal momenta of twisted states. In order to illustrate the difference between the scattering of twisted and plane--wave beams for both, single--molecule and macroscopic--target scenarios, detailed calculations have been performed for a H$_2$ target.      
\end{abstract}

\maketitle

%
%
\section{Introduction}

Only a decade after their theoretical prediction by Bliokh and co--workers \cite{BlB07} and experimental realization in electron microscopes \cite{UcT10,VeT10,McA11} twisted (or vortex) electron beams are now in the focus of intense research. These beams posses a helical phase front ${\rm e}^{i m \varphi}$ with $\varphi$ being the azimuthal angle about their propagation axis and $m$ the projection of the orbital angular momentum (OAM) upon this axis. In contrast to usual plane--wave states, for which $m = 0$, the OAM projection of twisted electrons can be as high as $\hbar m = 1000 \hbar$ \cite{MaT17}. Such a rather huge magnetic moment $\mu \propto m \mu_B$, where $\mu_B$ is the Bohr magneton, results from such an enormous $m$ and makes vortex beams particularly suitable for probing magnetic properties of materials at nano--scale \cite{GrH17,Bli17,LlB17}. The OAM--induced moment $\mu$ also allows to significantly enhance and explore magnetic phenomena in  electron--light coupling \cite{GaM12,BlS12,IvK13}. Moreover, a non--zero orbital angular momentum projection $m$, carried by vortex beams, can significantly influence fundamental atomic \textit{collision processes}. In particular, a large number of studies have been recently reported on OAM--effects in elastic scattering of vortex electrons by atomic and ionic targets \cite{VaB14,SeI15,KaK15,KaK17,KoZ18}. It was argued that the angular distribution and polarization of scattered electrons can be sensitive to kinematic parameters and projection of angular momentum $m$ of the incident beam. 

In contrast to elastic collisions with atoms, much less attention has been paid so far to the scattering of twisted electrons by \textit{molecules}. Electron--molecule scattering can provide, however, even  deeper insight into the properties of OAM beams than atomic studies since for  a molecular target the incident vortex electron interacts \textit{coherently} with two (or more) atomic centers that are spatially slightly displaced from each other. These centers act similar to the optical slits in the well--known Young's diffraction experiment, leading to an \textit{interference pattern} in the angular distribution of outgoing electrons. This Young--type interference is expected to be very sensitive to the complex internal structure of OAM beams. In particular, non--homogeneous phase and intensity profiles of twisted electron states can be probed by scattering off molecular targets. 

In this paper, we study theoretically the elastic scattering of twisted electrons by molecules. Special attention is paid to the question of how the (well--known) Young--type interference pattern in the angular distribution of outgoing electrons is influenced by the properties of OAM beams. To elucidate the main features of this OAM--dependence we employ the non--relativistic first Born approximation. This approach has been frequently applied and will be briefly recalled in Sec.~\ref{subsec:theory_plane_wave} for the scattering of plane--wave electrons by atomic and molecular targets. For the latter case we restrict ourselves to the simplest case of diatomic molecules whose structure is described by the independent atom model. We make use of this model to investigate the scattering of twisted electrons, as well. The wave function of these electrons, prepared in the Bessel state, is employed in Sec.~\ref{subsec:theory_twisted_wave} to derive the (first Born) scattering amplitude. With the help of this ``twisted'' amplitude we then derive the angular distribution of the outgoing electrons. Two scenarios are considered for our angle--resolved studies in which the Bessel beam collides with either (i) a single well--localized molecule or (ii) a macroscopic ensemble of randomly distributed molecules. Detailed calculations are performed for both scenarios and the H$_2$ target, and are discussed in Sec.~\ref{sec:results_discussion}. In particular, we show that the strength and position of interference minima and maxima in the angular distribution of electrons, scattered off a single molecule, may reflect both the phase and intensity structure of the incident twisted beam. For the macroscopic molecular target, the angular--differential scattering cross section depends on the (ratio of) transverse and longitudinal linear momenta of the twisted electrons. We finally summarize these results in Sec.~\ref{sec:Summary}.

Hartree atomic units ($\hbar = e = m_e = 1$) are used throughout the paper unless stated otherwise.

%
%
\section{Theoretical background}
\label{sec:theory}

\subsection{Scattering of plane--wave electrons}
\label{subsec:theory_plane_wave}

Not much has to be said about the potential scattering of plane--wave electrons by a single atom. Within the non--relativistic Born approximation, the theoretical analysis of this scattering is traced back to the amplitude
\begin{equation}
   \label{eq:scattering_amplitude_plane_wave}
   f_{1}^{\rm (pl)}({\bm k}, {\bm k}') = -\frac{1}{2 \pi} \, \int {\rm e}^{i ({\bm k} - {\bm k}'){\bm r}} \, V({\bm r}) \, {\rm d}{\bm r} \, .
\end{equation}
Here, ${\bm k}$ and ${\bm k}'$ are the momenta of the incident and outgoing electrons and $V({\bm r})$ is the potential that describes an electron--atom interaction. In the theory of atomic collisions $V({\bm r})$ is often expressed in terms of one or several Yukawa potentials
\begin{equation}
   \label{eq:Yukawa_potential}
   V({\bm r}) = - \frac{Z}{r} \, A \, {\rm e}^{-r d} \, ,
\end{equation}
that approximate the Coulomb field of the nucleus, screened by the target electrons. By inserting the potential (\ref{eq:Yukawa_potential}) into Eq.~(\ref{eq:scattering_amplitude_plane_wave}) one obtains the well--known expression for the Yukawa scattering amplitude:
\begin{equation}
   \label{eq:scattering_amplitude_plane_wave_Yukawa}
   f_{1}^{\rm (pl)}({\bm k}, {\bm k}') = \frac{2 Z A}{d^2 + q^2} \, ,
\end{equation}
where ${\bm q} = {\bm k} - {\bm k}'$ is the momentum transfer, whose square is given by $q^2 = 4 k^2 \sin^2\left(\theta/2\right)$. Here we made use of the fact that in elastic scattering both, the initial and final electron momenta have the same absolute value, i.e. $k = \left|{\bm k} \right| = \left| {\bm k}' \right|$, and $\cos \theta = \left({\bm k} {\bm k}' \right) / \left(k k' \right)$. We can use the amplitude (\ref{eq:scattering_amplitude_plane_wave_Yukawa}) to obtain the angle--differential  cross section:
\begin{eqnarray}
   \label{eq:cross_sections_plane_wave}
   \frac{{\rm d}\sigma^{\rm (pl)}_1}{{\rm d}\Omega} &=& \left| f_{1}^{\rm (pl)}({\bm k}, {\bm k}') \right|^2 \nonumber \\
   &=& \frac{4 Z^2 A^2}{\left(d^2 + 4 k^2 \sin^2\left(\theta/2\right) \right)^2} \, .  
\end{eqnarray}
for a Yukawa potential (\ref{eq:Yukawa_potential}). 

The Born approximation can also be used to describe---at least approximately---the electron scattering by not only atomic but also \textit{molecular} targets. For collisions between electrons and a neutral diatomic molecule, the essential features of the scattering process can be understood within the so--called independent atom model \cite{WiW77,MiM12}. Within this model, the net scattering amplitude
\begin{eqnarray}
   \label{eq:scattering_amplitude_plane_wave_molecule}
   f_{2}^{\rm (pl)}({\bm k}, {\bm k}') &=& -\frac{1}{2 \pi} \, \int {\rm e}^{i ({\bm k} - {\bm k}'){\bm r}} \, V({\bm r}) \, {\rm d}{\bm r} \nonumber \\
   &-& \frac{1}{2 \pi} \, \int {\rm e}^{i ({\bm k} - {\bm k}'){\bm r}} \, V({\bm r} + {\bm R}) \, {\rm d}{\bm r} \, \nonumber \\[0.2cm]
   &=& f_{1}^{\rm (pl)}({\bm k}, {\bm k}') \, \left(1 + {\rm e}^{-i {\bm q}{\bm R}} \right)
\end{eqnarray}
is expressed as a sum of amplitudes from two independent molecular centers, whose displacement with respect to each other is decribed by the vector ${\bm R}$. After some simple algebra one find, based on this expression, the angle--differential cross section:
\begin{eqnarray}
   \label{eq:cross_sections_plane_wave_molecule}
   \frac{{\rm d}\sigma^{\rm (pl)}_2}{{\rm d}\Omega} &=& \left| f_{2}^{\rm (pl)}({\bm k}, {\bm k}') \right|^2 \nonumber \\
   &=& 4 \cos^2\left({\bm q}{\bm R}/2\right) \, \frac{{\rm d}\sigma^{\rm (pl)}_1}{{\rm d}\Omega} \, .
\end{eqnarray}
Here, ${\rm d}\sigma^{\rm (pl)}_1 / {\rm d}\Omega$ is the single--center cross section (\ref{eq:cross_sections_plane_wave}) and the cosine squared term arises due to the interference between the electron waves emerging from both scattering centers. 

\subsection{Scattering of twisted electrons electrons}
\label{subsec:theory_twisted_wave}

\subsubsection{Monochromatic Bessel electrons}
\label{subsubsec:Bessel_electron_definition}

After having recalled the basic expressions for the potential scattering of plane--wave electrons, we are ready to consider the case of an incident \textit{twisted} beam. In order to start with this case we first need to agree on how to describe twisted electrons. In the present work we will assume that electrons are prepared in a state with well--defined energy $\varepsilon$, longitudinal momentum $k_z$, and projection $m$ of the orbital angular momentum onto the quantization (z--) axis. These so--called Bessel solutions of the field--free Schr\"odinger equation are described by the wave function
\begin{eqnarray}
	\label{eq:Bessel_electrons_definition}
	\psi^{(tw)}_{\varkappa m}({\bm r}) = \int \frac{{\rm d}^2{\bm k}_{\perp}}{\left(2 \pi\right)^2}
	\, a_{\varkappa m}({\bm k}_{\perp}) \, {\rm e}^{i {\bm k}{\bm r}} \, ,
\end{eqnarray}
with the amplitude 
\begin{equation}
	\label{eq:a_amplitued}
	a_{\varkappa m}({\bm k}_{\perp}) = (-i)^m \, \frac{1}{\sqrt{2 \pi \varkappa}} \, 
	{\rm e}^{i m \varphi_k} \, \delta\left(\left|{\bm k}_\perp\right| - \varkappa \right) \, ,
\end{equation}
and where the absolute value of the transverse momentum is fixed to $\varkappa = \sqrt{2 \varepsilon - k_z^2}$. A seen from these expressions, the Bessel electron state can be understood as a coherent superposition of the plane waves ${\rm e}^{i {\bm k} {\bm r}}$, whose wave vectors 
\begin{equation}
	\label{eq:wave_vectors_twisted}
    {\bm k} = \left({\bm k}_\perp, k_z \right) = \left(\varkappa \cos\varphi_k, \varkappa \sin\varphi_k, k_z \right)
\end{equation}
lay on the surface of a cone with the \textit{opening angle} $\tan\theta_k = \varkappa/k_z$. During the recent years Bessel beams of electrons have been discussed in a number of theoretical works \cite{MaH14,KaK17,Bli17}. We refer to these publications for all further details about the properties of Bessel electron states.

\subsubsection{Scattering amplitudes}
\label{subsubsec:Bessel_electron_amplitudes}
 
By making use of the Bessel electron wavefunction $\psi^{(tw)}_{\varkappa m}({\bm r})$ we can evaluate now the corresponding scattering amplitude. Similar to the plane--wave case we start our analysis with the potential scattering by a single atom. Within the Born approximation the amplitude reads:
\begin{eqnarray}
	\label{eq:scattering_amplitude_twisted}
	f_{1}^{\rm (tw)}(\varkappa, k_z , {\bm k}') && \nonumber \\[0.2cm]
	&& \hspace{-1.5cm} = -\frac{1}{2 \pi} \, 
	\int {\rm e}^{-i {\bm k}' {\bm r}} \, V({\bm r}+{\bm b}) \, \psi^{(tw)}_{\varkappa m}({\bm r}) \, {\rm d}{\bm r} \, ,
\end{eqnarray}
where the incident electron wavefunction $\psi^{(tw)}_{\varkappa m}({\bm r})$ is given by Eq.~(\ref{eq:Bessel_electrons_definition}) while the final state is described by a plane wave ${\rm e}^{i {\bm k}' {\bm r}}$, again. Thus, we assume that the outgoing electrons are \textit{observed} by conventional plane--wave detectors that are sensitive to their wave--vector ${\bm k}'$ only.  

In Eq.~(\ref{eq:scattering_amplitude_twisted}), moreover, we introduced the \textit{impact parameter} ${\bm b} = \left(b_x, b_y, 0 \right) = \left(b \cos\varphi_b, b \sin\varphi_b, 0 \right)$ in order to specify the lateral position of an atom within the incident electron wave--front. This parameter is essential since in contrast to a plane--wave, Bessel beams have a much more complex internal structure. In
particular, their intensity distribution in the plane transverse to the propagation direction, is not uniform but consists of concentric rings of high and low intensity. The direction of the local linear momentum also varies significantly within the wavefront. One therefore expects that the properties of the scattered electrons strongly depend on the position ${\bm b}$ of an atom with respect to the Bessel beam axis.

By inserting the incident wavefunction (\ref{eq:Bessel_electrons_definition}) into Eq.~(\ref{eq:scattering_amplitude_twisted}) we can express the ``twisted'' scattering amplitude in terms of the corresponding plane--wave amplitude:
\begin{eqnarray}
   \label{eq:scattering_amplitude_twisted_2}
   f_{1}^{\rm (tw)}(\varkappa, k_z , {\bm k}') && \nonumber \\[0.2cm]
   && \hspace*{-2cm} = -\frac{1}{2 \pi} \, 
   \int \frac{{\rm d}^2{\bm k}_{\perp}}{\left(2 \pi\right)^2}
   \, a_{\varkappa m}({\bm k}_{\perp})
   \int {\rm d}{\bm r} \, {\rm e}^{i {\bm q} {\bm r}} \, V({\bm r}+{\bm b}) \nonumber \\[0.2cm]
   && \hspace*{-2cm} = \int \frac{{\rm d}^2{\bm k}_{\perp}}{\left(2 \pi\right)^2}
   \, a_{\varkappa m}({\bm k}_{\perp}) \, {\rm e}^{- i {\bm q} {\bm b}} \, 
   f_{1}^{\rm (pl)}({\bm k}, {\bm k}') \, .
\end{eqnarray}
The integration over the \textit{absolute value} of the transverse momentum $k_\perp = \left| {\bm k}_\perp \right|$ can be easily carried out by employing the explicit form of the amplitude $a_{\varkappa m}({\bm k}_{\perp})$:
\begin{eqnarray}
   \label{eq:scattering_amplitude_twisted_3}
   f_{1}^{\rm (tw)}(\varkappa, k_z , {\bm k}') &=&
   \frac{(-i)^m}{2 \pi} \, \sqrt{\frac{\varkappa}{2 \pi}} \, {\rm e}^{i {\bm k}' {\bm b}} \nonumber \\[0.2cm]
   && \hspace*{-1.5cm} \times \int \frac{{\rm d}\varphi_k}{2 \pi} \, {\rm e}^{i m \varphi_k - i {\bm k}_{\perp} {\bm b}} 
   \, f_{1}^{\rm (pl)}({\bm k}, {\bm k}') \, ,
\end{eqnarray}
where ${\bm k}_\perp {\bm b} = \varkappa b \cos\left(\varphi_k - \varphi_b \right)$ and the incident electron wave--vector in the plane--wave amplitude is given by Eq.~(\ref{eq:wave_vectors_twisted}). Further evaluation of the $f_{1}^{\rm (tw)}(\varkappa, k_z , {\bm k}')$ requires the knowledge about the explicit form of the scattering potential $V({\bm r})$ and, hence, the plane--wave amplitude (\ref{eq:scattering_amplitude_plane_wave}). For example, the integral in Eq.~(\ref{eq:scattering_amplitude_twisted_3}) has to be calculated numerically if Yukawa potential (\ref{eq:Yukawa_potential}) is used to describe the electron--atom interaction \cite{KaK17,SeI15}.

\begin{figure}
	\centering
	\includegraphics[width=0.8\linewidth]{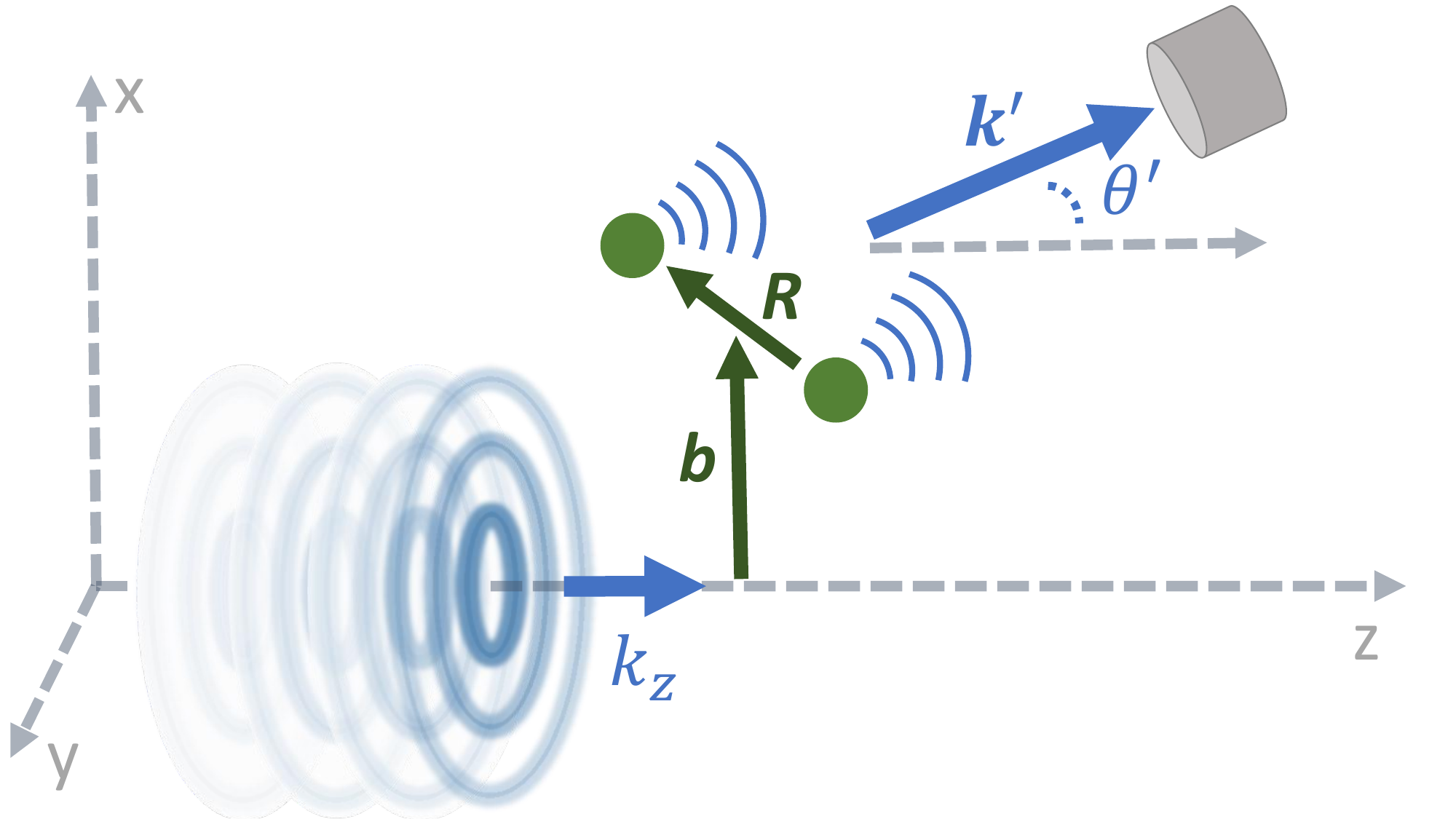}
	\caption{(Color online) The geometry of the potential scattering of twisted electrons by a diatomic molecule. The beam propagation direction is chosen as the $z$--axis. Together with the molecular internuclear vector ${\bm R}$ this axis defines the $xz$--plane. The intensity distribution of Bessel electrons in the $xy$--plane, normal to the propagation direction, has a concentric ring structure with a central dark spot at the impact parameter ${\bm b} = 0$. Finally, the scattered electrons are detected in the plane of molecule ($xz$--plane).}
	\label{Fig1}
\end{figure}

The theory of potential scattering of Bessel electrons can be easily generalized towards diatomic molecular targets. Similar to the plane--wave case we employ the independent atom model and write the scattering amplitude as:
\begin{eqnarray}
	\label{eq:scattering_amplitude_twisted_molecule}
	f_{2}^{\rm (tw)}(\varkappa, k_z , {\bm k}') &=& \nonumber \\[0.2cm]
	&& \hspace{-2.5cm} = -\frac{1}{2 \pi} \, 
	\int {\rm e}^{-i {\bm k}' {\bm r}} \, V({\bm r}+{\bm b}+{\bm R}/2) \, \psi^{(tw)}_{\varkappa m}({\bm r}) \, {\rm d}{\bm r} \nonumber \\[0.2cm]
	&& \hspace{-2.2cm} - \frac{1}{2 \pi} \, 
	\int {\rm e}^{-i {\bm k}' {\bm r}} \, V({\bm r}+{\bm b}-{\bm R}/2) \, \psi^{(tw)}_{\varkappa m}({\bm r}) \, {\rm d}{\bm r} \, .
\end{eqnarray}
Here, ${\bm R}$ is the internuclear vector and ${\bm b}$ is the position of the molecular center with respect to the central axis of the incident beam, see Fig.~\ref{Fig1}. Again, by using the explicit form of the wave--function $\psi^{(tw)}_{\varkappa m}({\bm r})$ we can further evauate the expression (\ref{eq:scattering_amplitude_twisted_molecule}) as:
\begin{eqnarray}
   \label{eq:scattering_amplitude_twisted_molecule_2}
   f_{2}^{\rm (tw)}(\varkappa, k_z , {\bm k}') &=& \nonumber \\[0.2cm]
   && \hspace{-2.5cm} = 2 \, \int \frac{{\rm d}^2{\bm k}_{\perp}}{\left(2 \pi\right)^2} \, a_{\varkappa m}({\bm k}_{\perp}) \, {\rm e}^{- i {\bm q} {\bm b}} \, \cos\left({\bm q}{\bm R}/2\right) \,
   f_{1}^{\rm (pl)}({\bm k}, {\bm k}') \nonumber \\[0.2cm]
   && \hspace{-2.5cm} = \frac{(-i)^m}{\pi} \, \sqrt{\frac{\varkappa}{2 \pi}} \, {\rm e}^{i {\bm k}' {\bm b}} \nonumber \\[0.2cm]
   && \hspace*{-2.5cm} \times \int \frac{{\rm d}\varphi_k}{2 \pi} \, {\rm e}^{i m \varphi_k - i {\bm k}_{\perp} {\bm b}} \, \cos\left({\bm q}{\bm R}/2\right)
   \, f_{1}^{\rm (pl)}({\bm k}, {\bm k}') \, ,
\end{eqnarray}
where the plane--wave scattering amplitude $f_{1}^{\rm (pl)}({\bm k}, {\bm k}_f)$ is given by Eq.~(\ref{eq:scattering_amplitude_plane_wave_Yukawa}) and the integration over the azimuthal angle $\varphi_k$ is performed numerically.

\subsubsection{Angular distribution of scattered electrons}
\label{subsubsec:Bessel_electron_angular distribution}

Since the scattering amplitude (\ref{eq:scattering_amplitude_twisted_molecule_2}) explicitly depends on the impact parameter $\bm{b}$ and the internuclear vector $\bm{R}$, we need first to discuss the composition and localization of the molecular target. In our present study we will discuss two scenarious in which the target consists of (i) a single molecule located at a well--defined position in the incident beam, and (ii) molecules, randomly distributed over the entire volume. The angular distribution of scattered electrons for the first case can be written as:
\begin{eqnarray}
	\label{eq:angular_distribution_single_molecule}
	W_2^{({\rm tw})}(\vartheta', \varphi'; \, {\bm b}, m) &=& {\cal N} \left| f_{2}^{\rm (tw)}(\varkappa, k_z , {\bm k}') \right|^2 \, ,
\end{eqnarray}
where the amplitude $f_{2}^{\rm (tw)}(\varkappa, k_z , {\bm k}')$ is given by Eq.~(\ref{eq:scattering_amplitude_twisted_molecule_2}) and the prefactor ${\cal N}$ is defined by the normalization condition $\int {\rm d}\Omega' \left|W^{({\rm tw})}(\vartheta', \varphi'; \, {\bm b}, m)\right|^2 = 1$. Moreover, in this expression $\vartheta'$ and $\varphi'$, are the polar and azimuthal angles of the scattered electron wave--vector ${\bm k}'$. In the calculations below we will assume that the outgoing electrons are detected in the $xz$--plane, spanned by the beam axis and the internuclear vector ${\bm R}$ of the molecule. For this choice of geometry the azimuthal angle is $\varphi' = 0$.

While the scattering of twisted electrons by a \textit{single} molecule might be used to obtain detailed information about the properties of Bessel beams, this scenario can hardly be realized in current experiments. Therefore, we will consider yet another case in which Bessel electrons collide with a \textit{macroscopic} molecular target. This target can be described as an ensemble of aligned molecules that are randomly and uniformly distributed over the transverse extent of the incident beam. The angle--differential cross section for such a target,
\begin{eqnarray}
	\label{eq:differential_cross_section_molecule}
	\frac{{\rm d}\sigma^{({\rm tw})}_2}{{\rm d}\Omega'}(\vartheta', \varphi') &=& \frac{1}{\cos\theta_k}
	\, \int \left| f_{2}^{\rm (tw)}(\varkappa, k_z , {\bm k}') \right|^2 {\rm d}^2{\bm b} \nonumber \\[0.2cm]
	&& \hspace*{-2cm} = \frac{4}{\cos\theta_k} \int \frac{{\rm d}\varphi_k}{2 \pi} \,
	\cos^2\left({\bm q} {\bm R}/ 2 \right) \, \left| f_{1}^{\rm (pl)}({\bm k}, {\bm k}')\right|^2 \, ,
\end{eqnarray}
can be obtained upon averaging the square of the transition amplitude (\ref{eq:scattering_amplitude_twisted_molecule_2}) over the impact parameters ${\bm b}$. The pre--factor, moreover, has been derived by normalizing the result by the number of incident electrons, see Refs.~\cite{SeI15,KaK15} for further details. As seen from the second line of Eq.~(\ref{eq:differential_cross_section_molecule}), the angle--differential cross section ${\rm d}\sigma^{({\rm tw})}_2/{\rm d}\Omega'$ for a macroscopic target depends neither on the projection $m$ of the orbital angular momentum nor on the spatial structure of an incident phase front. It is still sensitive, however, to the ratio of the transverse to the longitudinal momenta of electrons as described by the opening angle $\theta_k$.

%
%
\section{Computational details}
\label{sec:computational_details}

For the analysis of the effects introduced by twisted electron beams instead of plane waves we choose molecular hydrogen H$_2$ as a relatively simple test target. We describe this molecule as two independent hydrogen atoms, displaced from each other by the dirtance $R$. In our calculations, $R = 1.401$~a.u. is the most probable internuclear distance and the electrostatic potential of (individual) hydrogen atoms is approximated by a sum of two Yukawa potentials:
\begin{equation}
	\label{eq:hydrogen_potential}
	V_{H}(r) = -\frac{e^2}{r} \, \sum\limits_{i=1,2} A_i {\rm e}^{-r d_i} \, ,
\end{equation}
where the parameters $A_i$ and $d_i$ are determined by a fit to the results of Dirac--Hartree--Fock--Slater (DHFS) self--consistent calculations \cite{SaM87,Sal91}. In this approximation, the single--center plane--wave amplitude $f_{1}^{\rm (pl)}({\bm k}, {\bm k}')$ is expressed analytically by a sum of Yukawa terms (\ref{eq:scattering_amplitude_plane_wave_Yukawa}). Upon inserting this amplitude into Eqs.~(\ref{eq:scattering_amplitude_twisted_molecule_2}) and (\ref{eq:differential_cross_section_molecule}) the integration over the azimuthal angle $\varphi_k$ is performed numerically by means of Gauss--Legendre  quadrature.

%
%
\section{Results and discussion}
\label{sec:results_discussion}

\subsection{Scattering by a well--localized H$_2$ molecule}
\label{subsec:Results_single_molecule}

\begin{figure}
	\centering
	\includegraphics[width=0.9\linewidth]{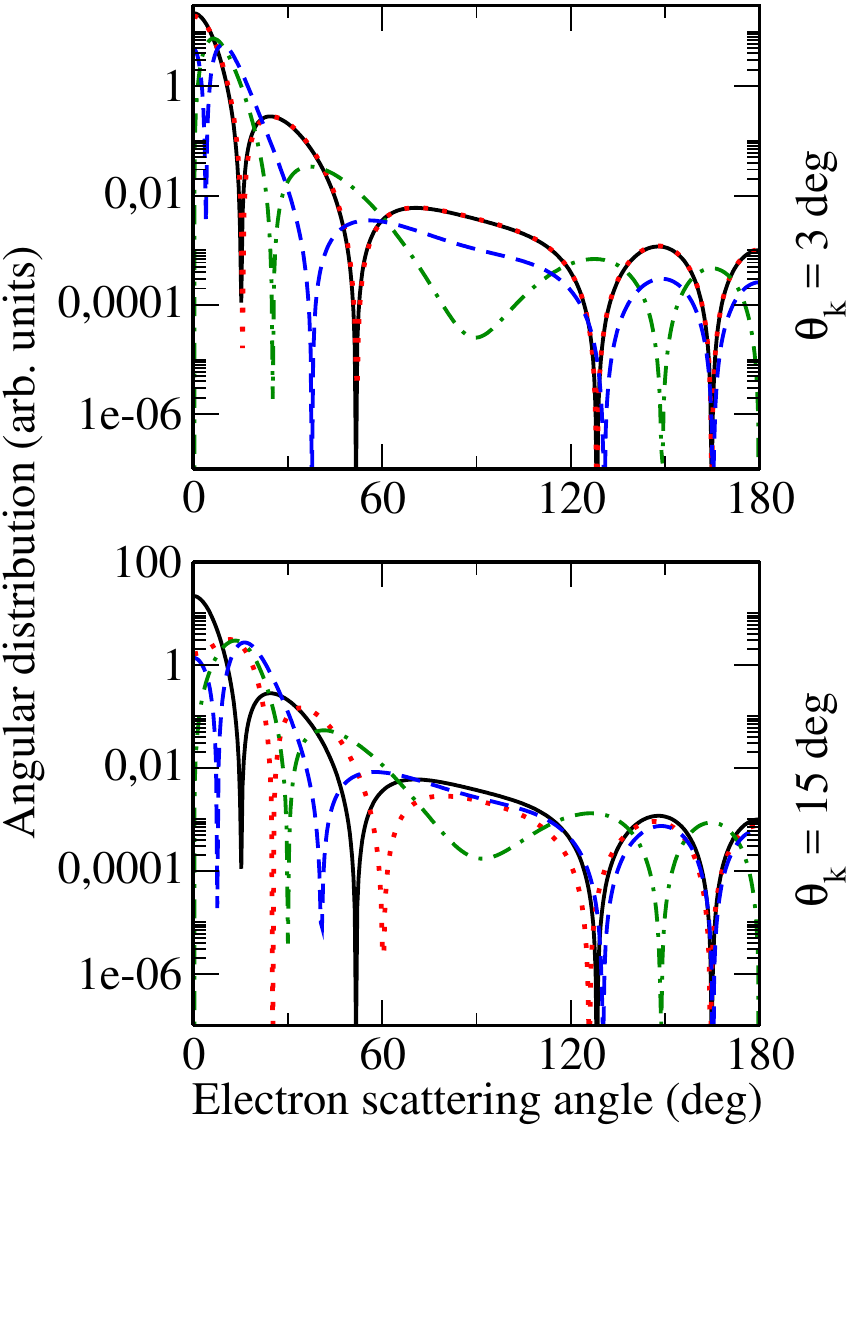}
	\vspace*{-2cm}
	\caption{(Color online) The angular distribution of electrons scattered by a single H$_2$ molecule. Calculations have been performed for an incident Bessel beam with kinetic energy 1 keV, opening angles $\theta_k$ = 3~deg (top panel) and $\theta_k$ = 15~deg (bottom panel), and OAM projections $m$ = 0 (red dotted line), $m$ = 1 (green dash--dotted line), and $m$ = 2 (blue dashed line). Moreover, the target molecule is oriented perpendicular to the beam axis and its center of mass is placed at this axis. The results of the calculations are compared with the prediction obtained for an incident plane--wave electrons (solid black curve).}
	\label{Fig2}
\end{figure}

We mentioned already above that the analysis of the potential scattering of twisted electrons by diatomic molecules requires the knoweledge about the composition of the target. In Section~\ref{subsubsec:Bessel_electron_angular distribution} we have discussed two cases in which the Bessel beam collides with either (i) a single molecule, or (ii) a macroscopic molecular target. Even though the second scenario is much more feasible with current experimental techniques, we discuss the scattering of electrons by a (single) well--localized molecule first. This single--scatterer scenario will allow us to better understand the properties of twisted electrons and the details of their interaction with diatomic targets. As seen from Eq.~(\ref{eq:angular_distribution_single_molecule}), the angular distribution of outgoing electrons depends in this case both, on the OAM projection $m$ of the beam and on the impact parameter ${\bm b}$ of the molecule. In Fig.~\ref{Fig2}, for example, we display the scattering patterns for a H$_2$ molecule, oriented perpendicular to the incident beam axis, ${\bm R} \perp z$, and whose center of mass is placed at this axis, ${\bm b} = 0$. Calculations have been performed for two opening angles of the incident Bessel electrons, $\theta_k$~=~3~deg (top panel) and $\theta_k$~=~15~deg (bottom panel), and for three projections of their orbital momentum, $m$~=~0 (red dotted line), $m$~=~1 (green dash--dotted line) and $m$~=~2 (blue dashed line). The angular distributions $W_2^{({\rm tw})}(\vartheta', \varphi'; \, {\bm b}, m)$ are also compared here with those for plane--wave (black solid line). As seen from the figure, the scattering of plane--wave and Bessel electrons with $zero$ OAM projection and small opening angle, $\theta_k =$~3~deg, yield almost identical angular distributions. This result is well expected since the Bessel electron wave function (\ref{eq:Bessel_electrons_definition}) recovers, for $m$~=~0 and $\theta_k \, \to$~0, the standard solution for a plane wave that propagates along the $z$ axis, see Ref.~\cite{SeI15} for further details. 

For non--zero values of $m$ the scattering pattern of twisted electrons may differ from the plane--wave predictions even for very small opening angles. For example, as seen in the top panel of Fig.~\ref{Fig2}, the positions of maxima and minima of the angular distribution $W_2^{({\rm tw})}(\vartheta', 0; \, {\bm b} = 0, m = 1)$ for $m$~=~1 are almost inverted compared to the plane--wave case. In order to explain this behaviour we shall recall again that the oscillations in the angle--differential cross sections arise from the interference between electrons scattered by two molecular centers. In classical wave mechanics this interference can be understood based on the analysis of the phase shift $\Delta\phi$ between the electron waves. For the plane--wave incident electrons, for example, the well known expression
\begin{equation}
	\label{eq:phase_difference_plane_wave}
	\Delta\phi^{({\rm pl})} = k' R \sin\theta' \, 
\end{equation}
is trivially derived from the path length difference $\Delta r = R \sin\theta'$ between two waves travelling at the angle $\theta'$ with respect to the $z$--axis. Here we assumed that these waves are originated from two point--like scatterers, located at the distance $R$ from each other, and where the internuclear vector ${\bm R}$ is normal to the beam direction. With the help of the phase difference (\ref{eq:phase_difference_plane_wave}) one can easily estimate the oscillatory behaviour of the electron angular distribution:
\begin{eqnarray}
	\label{eq:cross_sections_plane_wave_estimate}
	W^{({\rm pl})}_2(\theta') &\propto& \left| {\rm e}^{i {\bm k}' {\bm r}} + {\rm e}^{i {\bm k}' {\bm r} + i \Delta\phi^{({\rm pl})}} \right|^2 \nonumber \\[0.2cm]
	&=& 4 \cos^2\left(\frac{k' R \sin\theta'}{2} \right) \, .
\end{eqnarray}
This is exactly the interference term from Eq.~(\ref{eq:cross_sections_plane_wave_molecule}) for $z \perp {\bm R}$ and a plane wave travelling in $z$ direction. In this case we have ${\bm q}{\bm R}/2 = - {\bm k}'{\bm R}/2 = - k' R \sin\theta'$. 

In contrast to a plane--wave, twisted electrons already carry by themselves an \textit{additional} phase shift at molecular centers. This shift arises from the helical structure of the Bessel phase front. That is, the wave--function (\ref{eq:Bessel_electrons_definition}) of incident twisted electrons can be written---upon the integration over the transverse momentum ${\bm k}_\perp$---as:
\begin{equation}
	\label{eq:Bessel_electron_wavefunction_coordinate}
	\psi^{(tw)}_{\varkappa m}({\bm r}) = \frac{{\rm e}^{i m \varphi_r}}{\sqrt{2 \pi}} \,
	\sqrt{\varkappa} \, {\rm e}^{i k_z z} \, J_m\left( \varkappa r_\perp \right) \, ,
\end{equation}
where $J_m$ is the Bessel function and $\varphi_r$ is the azimuthal angle of the position vector ${\bm r}$. As seen from this expression, the phase of the incident twisted electrons at the positions of the two scatterers, $+ {\bm R}/2$ and $- {\bm R}/2$, differs by $m \pi$. This phase shift should be added to the standard term $k' R \sin\theta'$ that arises from the difference of paths of outgoing electrons:
\begin{equation}
\label{eq:phase_difference_twisted_electrons}
\Delta\phi^{({\rm tw})} = \pi m + k' R \sin\theta' \, .
\end{equation}
This expression, derived again for point--like scatterers, allows for a qualitative understanding of the oscillatory behaviour of the angular distribution (\ref{eq:angular_distribution_single_molecule}) of scattered electrons:
\begin{eqnarray}
	\label{eq:angular_distribution_single_molecule_etimate}
	W_2^{({\rm tw})}(\vartheta', 0; \, 0, m) &\propto&  4 \cos^2\left(\frac{k' R \sin\theta'}{2} + \frac{\pi m}{2} \right) \, ,
\end{eqnarray}
for the case of an incident Bessel beam. 

By comparing Eqs.~(\ref{eq:cross_sections_plane_wave_estimate}) and (\ref{eq:angular_distribution_single_molecule_etimate}) we see that the phase of the interference pattern for a Bessel beam with \textit{odd} OAM projections is shifted by $\pi/2$ with respect to that of plane--wave electrons. This leads to the inversion of maxima and minima in the electron emission spectra as displayed in Fig.~\ref{Fig2}. In contrast, the angular distribution of scattered electrons from incident Bessel beams with \textit{even} $m$ should resemble the plane--wave result (\ref{eq:cross_sections_plane_wave_molecule}). Our calculations for $m = 2$ (blue dashed line) confirm this prediction for large angles $\theta'$ but indicate that $W_2^{({\rm tw})}(\theta', 0; 0, m = 2)$ and $W_2^{({\rm pl})}$ might differ for small $\theta'$'s. This discrepancy between Bessel and plane--wave calculations can be easily understood by the fact that forward scattering involves a small momentum transfer. In this case an incident electron interacts mainly not with (rather small) nucleus but with valence target electrons. 
The molecular centers can not be treated as point--like scatterers and, hence, the approximate expression (\ref{eq:angular_distribution_single_molecule_etimate}) is not valid anymore.  

\begin{figure}
	\centering
	\includegraphics[width=0.9\linewidth]{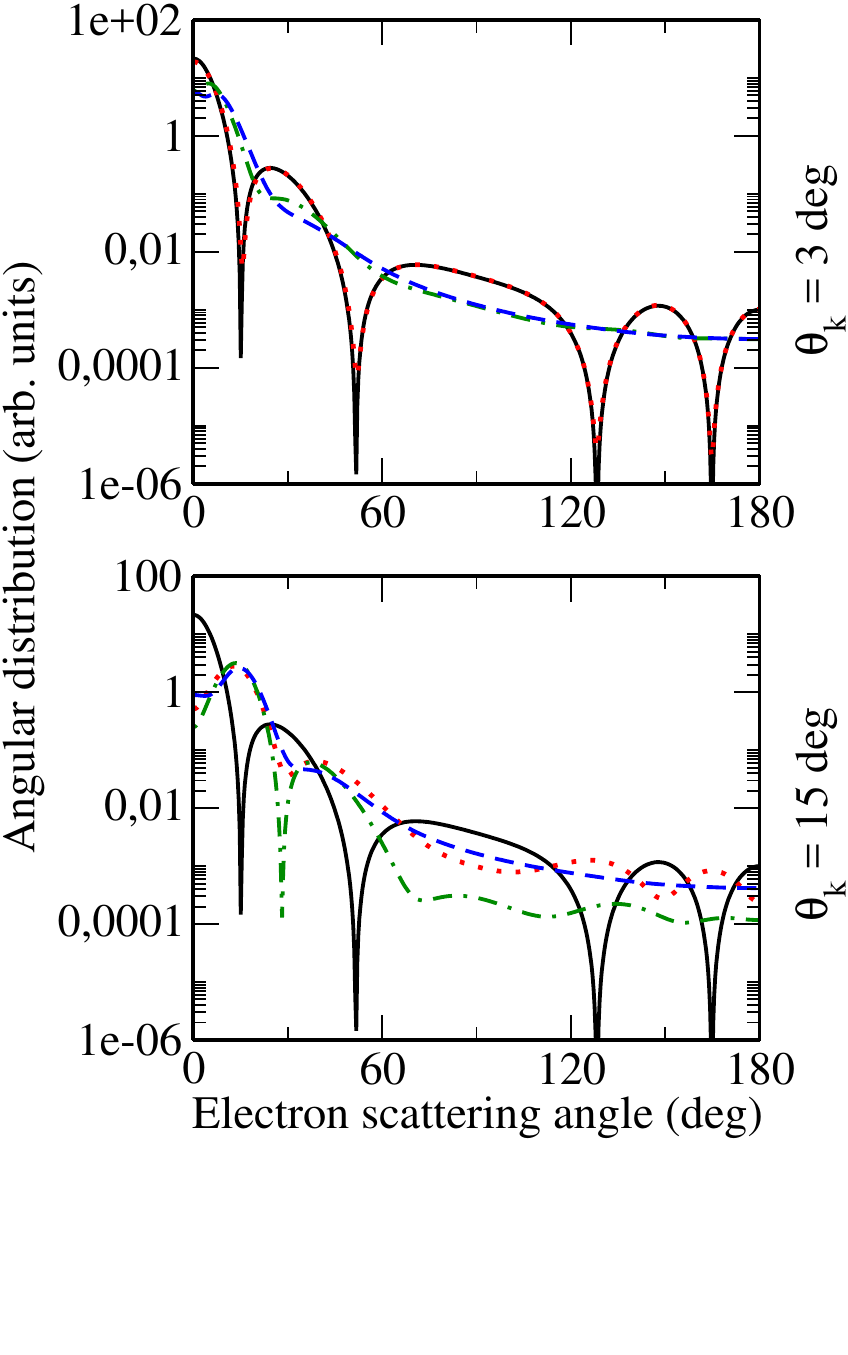}
	\vspace*{-2cm}
	\caption{(Color online) The same as Fig.~\ref{Fig2} but for the H$_2$ molecule located in such a way that one of its nuclei lays on the beam axis.}
	\label{Fig3}
\end{figure}

Until now we have discussed the scattering of twisted electrons by a single H$_2$ molecule whose center of mass is placed at the beam axis, ${\bm b}~=~0$. We have seen that this geometry allows one to explore the phase structure of the Bessel beam. In order to probe its \textit{intensity pattern}, one needs to shift the molecule in such a way that one atomic center is placed at the beam axis and the other one at ${\bm r} = {\bm R}$, respectively. Similar to before we will assume here that the internuclear vector is normal to the incident beam direction, ${\bm R} \perp z$. For these position and orientation of the molecule its atomic centers are exposed to a different intensity of the incident electrons,
\begin{equation}
	\label{eq:Bessel_beam_intensity}
	\rho^{(tw)}_{\varkappa m}({\bm r}) = \left| \psi^{(tw)}_{\varkappa m}({\bm r}) \right|^2
	= \frac{\varkappa}{(2 \pi)} \, J_m^2(\varkappa r_\perp) \, ,
\end{equation}
and, hence, the scattering can be understood as a molecular analogue of Young's experiment with two slits of unequal widths. 

\begin{figure}[t]
	\centering
	\includegraphics[width=0.85\linewidth]{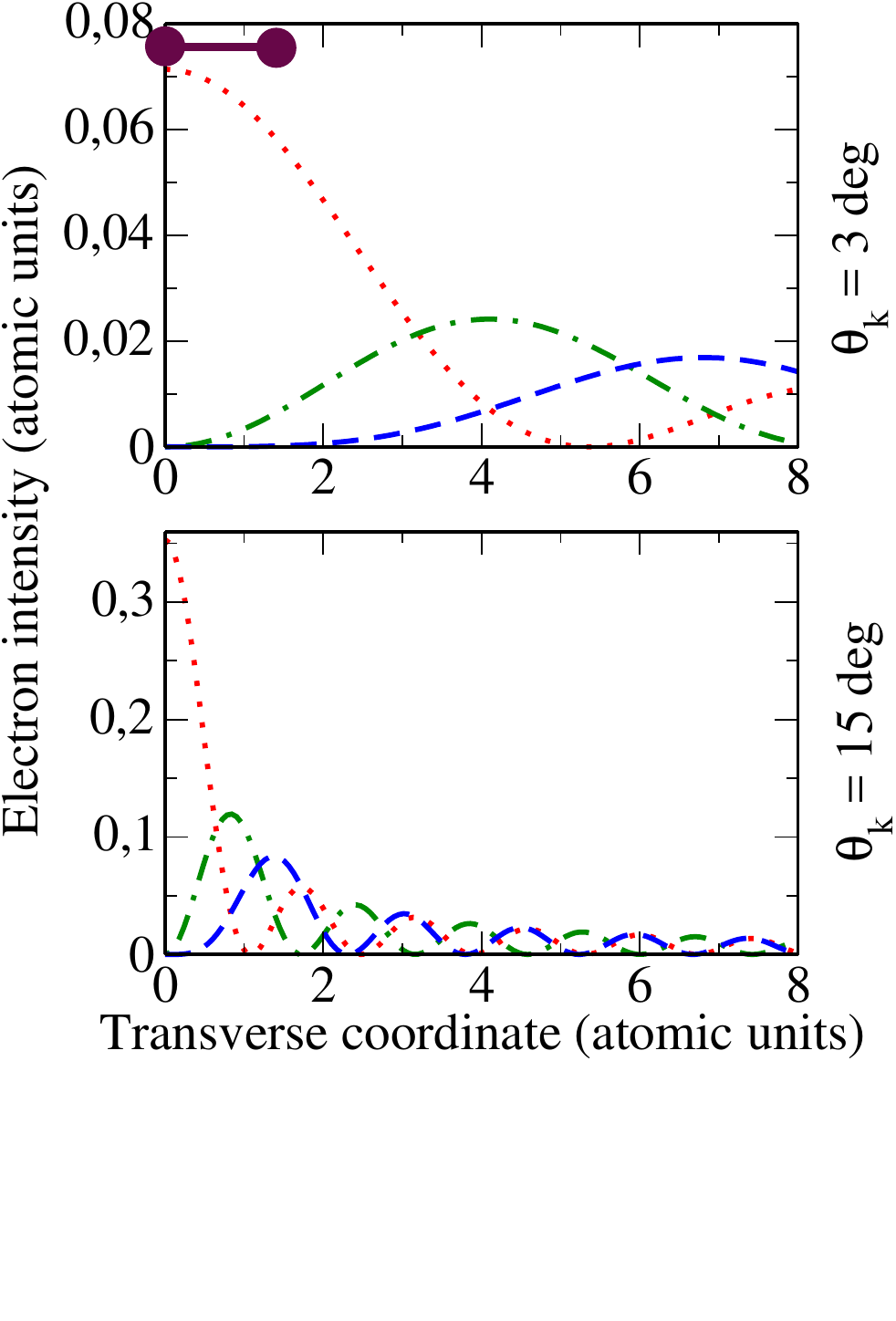}
	\vspace*{-2cm}
	\caption{(Color online) The transverse intensity (\ref{eq:Bessel_beam_intensity}) of the Bessel electron beam with the energy 1~keV, opening angles $\theta_k$~=~3~deg (top panel) and 15~deg (bottom panel), and OAM projections $m$~=~0 (red dotted line), $m$~=~1 (green dash--dotted line) and $m$~=~2 (blue dashed line). For comparison, we also display the size of the H$_2$ molecule.}
	\label{Fig4}
\end{figure}

In Fig.~\ref{Fig3} we display the angular distribution (\ref{eq:angular_distribution_single_molecule}) for the case when one of molecular centers is located at the beam axis. Again, calculations have been performed for Bessel electrons with energy 1~keV, opening angles $\theta_k$~=~3~deg and 15~deg, and OAM projections $m$~=~0, 1 and 2. As seen from the figure, the interference pattern in the angular distribution $W_2^{({\rm tw})}$ is very sensitive to the orbital momentum projection $m$. For the opening angle $\theta_k$~=~3~deg, for example, the oscillations in $W_2^{({\rm tw})}$ are very pronounced for the scattering of twisted electrons with the OAM projection $m$~=~0, while they disappear almost entirely if  $m$~=~1 and 2. This OAM--behaviour is caused by the inhomogenious intensity distribution of the Bessel beam. As seen from Eq.~(\ref{eq:Bessel_beam_intensity}) and Fig.~(\ref{Fig4}), the intensity $\rho^{(tw)}_{\varkappa m}({\bm r})$ of incident electrons, as ``seen'' at both atomic centers of the molecule, is approximately the same for $m$~=~0, thus leading to the pronounced interference picture. For $m$~=~1 and 2, in contrast, one of the nuclei is located in the dark spot on the beam axis and, hence, the interference pattern in the $W_2^{({\rm tw})}$ almost disappears like in the single--scatterer case.

%
%

The oscillatory behaviour of the angular distribution of scattered electrons may change significantly also with the variation of the opening angle $\theta_k$. For instance, for $\theta_k = 15$~deg the $W_2^{({\rm tw})}$ is rather monotonic for $m$~=~0 while, in contrast, strongly oscillates for $m$~=~1, see bottom panel Fig.~\ref{Fig3}. Again, this can be explained by the fact that the intensity of the incident beam $\rho^{(tw)}_{\varkappa m}({\bm r})$ at the positions of molecular centers, varies significantly with the $\theta_k$ and $m$.

\subsection{Scattering by the macroscopic target}
\label{subsec:Results_macroscopic_target}

\begin{figure}[t]
	\centering
	\includegraphics[width=1.0\linewidth]{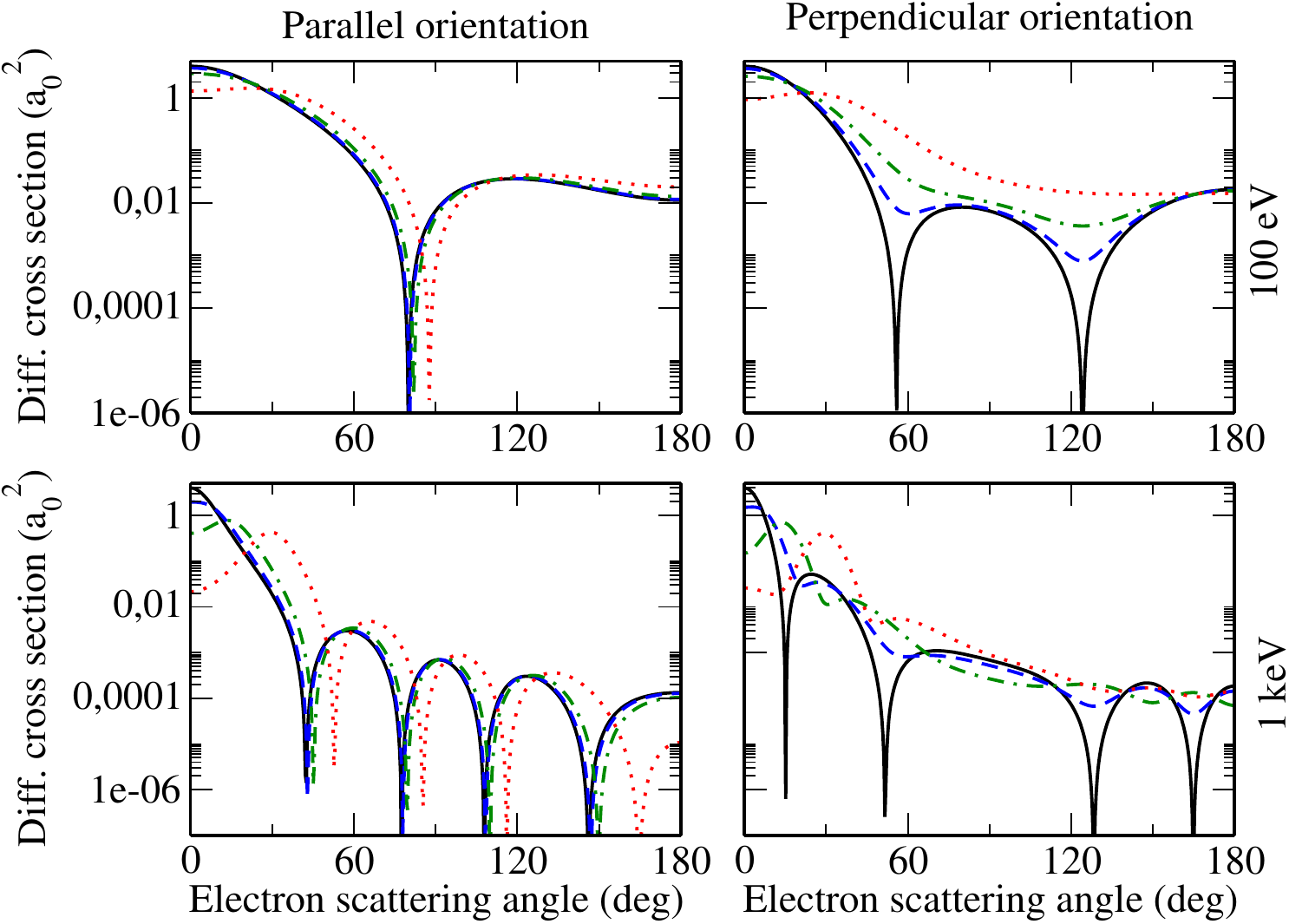}
	\caption{(Color online) Angle--differential cross section for the elastic scattering of electrons by a macroscopic hydrogenic target. Calculations have been performed for the incident electron energy 100 eV (top panel) and 1 keV (bottom panel) as well as for the alignment of H$_2$ molecules parallel (left column) and perpendicular (right column) to the beam axis. Predictions obtained for the Bessel beam with the opening angles $\theta_k$~=~7~deg (blue dashed line), 15~deg (green dash--dotted--line) and 30~deg (red dotted line) are compared with the plane--wave results (black solid line).}
	\label{Fig5}
\end{figure}

In the previous section we have discussed the scattering of Bessel electrons by a well--localized H$_2$ molecule. In a more realistic experimental scenario the twisted electron beam collides with a \textit{macroscopic} molecular target. We can describe such a target as an incoherent ensemble of aligned molecules that are randomly and homogeneously distributed over the cross sectional area of the electron beam. The scattering cross section (\ref{eq:differential_cross_section_molecule}), derived for this macroscopic case, is independent on the OAM projection but still sensitive to the kinematic parameters of the twisted beam as given by the opening angle $\theta_k$. In Fig.~\ref{Fig5} we display the ${\rm d}\sigma_2^{\rm (tw)}/{\rm d}\Omega'$ for three opening angles, $\theta_k$~=~7~deg, 15~deg and 30~deg, and compare our results with predictions for plane wave electrons (\ref{eq:cross_sections_plane_wave_molecule}). Calculations have been performed for kinetic electron energies 100~eV and 1~keV and for molecules aligned either parallel or perpendicular to the beam axis. 

%
%

Fig.~\ref{Fig5} shows that the angle--differential cross section (\ref{eq:cross_sections_plane_wave_molecule}) exhibits a qualitatively different behaviour if the molecular axis is oriented in parallel and perpendicular to beam axis. If these molecules are aligned along the incident beam direction (${\bm R} \, \parallel \, z$), ${\rm d}\sigma_2^{\rm (tw)}/{\rm d}\Omega'$ strongly oscillates as a function of scattering angle. The oscillatory pattern resembles very much that from the plane--wave electrons (black solid line) and is just \textit{shifted} towards larger $\theta'$'s. This shift of the angular distribution increases with the opening angle $\theta_k$; similar effect was previously predicted for the scattering of Bessel electrons by atomic targets \cite{SeI15,KoZ18}. It can be explained based on the standard representation of the Bessel state as a coherent sum of plane waves lying
on a momentum cone surface with the opening angle $\theta_k$. Since each plane--wave components ${\rm e}^{i {\bm k}{\bm r}}$ approaches the molecule not along the $z$--axis but under the angle $\theta_k \ne 0$ with respect to it, the minima and maxima of the angular distribution of scattered electrons are shifted by the angle $\Delta \theta' \approx \theta_k$, as seen in the left panels of Fig.~\ref{Fig5}.  

The oscillations in the differential cross section ${\rm d}\sigma_2^{\rm (tw)}/{\rm d}\Omega'$, clearly seen for the parallel (to the $z$--axis) alignment of molecules, are remarkably blurred if the H$_2$ dimers are oriented perpendicular to the incident beam; right column of Fig.~\ref{Fig5}. To explain this behaviour we recall that the characteristic size of the high-- and low--intensity rings in the incident electron intensity (\ref{eq:Bessel_beam_intensity}) is proportional to $r_\perp \propto 1/\varkappa = 1/\left(k \sin\theta_k \right)$. For relatively large opening angles, $\theta_k \gtrsim 5$~deg, and energies in the range from 100 eV to 1 keV this ring--like intensity varies over a spatial extent comparable to the size of the molecule, $R = 1.401$~a.u. Thus, if the molecules are aligned perpendicular to the beam direction their nuclei are often exposed to a different intensity of incident electron beam. As for a well--localized molecule, this leads to the decrease of the interference pattern, cf. Section~\ref{subsec:Results_single_molecule}. Similar effect was also reported recently for the photoionization of diatomic molecular ions by twisted light \cite{PeS15}. In contrast, for the parallel orientation both molecular centers always experience the same intensity of incident electrons and, hence, the oscillatory behaviour of the angle--differential cross section is well pronounced.

\section{Summary and outlook}
\label{sec:Summary}

In summary, we have presented a theoretical analysis of the elastic scattering of Bessel electrons by diatomic molecules. Special attention has been paid to the oscillatory behaviour of the angle--differential cross section, which arises from the interference due to the coherent interaction of electron with two atomic centers. In order to understand how this Young--type interference is influenced by the ``twistedness'' of the incident electron beam we have employed the independent atomic model to describe the molecule and the first Born approximation to evaluate the scattering amplitude. Within the framework of our theoretical model two scenarios have been discussed in which the twisted beam interacts either with (i) a single molecule or (ii) a macroscopic molecular target. In both cases detailed calculations have been performed for the hydrogen H$_2$ molecule.

The scenario in which twisted electrons scatter off a single molecule, even though being rather academic, allows one to better understand the properties of Bessel beams. For example, if the center of mass of the molecule is located at the beam axis the angular distribution of scattered electrons is very sensitive to the phase structure of the beam. The Young type minima and maxima in the electron scattering pattern are remarkably shifted due to the phase variation in Bessel beams with different OAM projection $m$. If, in contrast, not the molecular center but one of the nuclei is located at the beam axis the angular distribution reflects the inhomogeneous intensity profile of the Bessel state. A variation of $m$ can then either enhance or lower the interference of amplitudes from the (two) molecular centers, depending of whether they experience a similar or different intensity of incident electrons. 

For a macroscopic molecular target, in contrast, the angle--differential cross section is independent on the OAM projection $m$. However, the angle--differential cross section is very sensitive to the beam opening angle $\theta_k$ and the orientation of the molecules. For the molecules being aligned along the beam axis the angular distribution of scattered electrons exhibits an oscillatory behaviour very similar to that observed for the plane--wave electrons. The effect of the ``twistedness'' here is a shift of the minimum and maximum positions which becomes more pronounced as the opening $\theta_k$ increases. If the molecules are oriented perpendicular to the beam axis, the oscillatory behaviour is suppressed for all $\theta_k$ which is again explained by the inhomogeneous probability profile of the Bessel beam.

As seen from the results of the present study, the interference pattern in the angular distribution of scattered electrons can be very sensitive to both the OAM projection of the incident beam and to the geometry of the target. We expect that such a sensitivity may be present for the collisions of twisted electrons by complex molecules, as well. Of special interest here is the scattering off chiral molecules whose analysis is currently underway and will be presented in a separate paper.

\section*{Acknowledgements}

We like to thank Lea Schulze for discussions. A.V.M. acknowledges support from the BASIS foundation (Grant No. 17--13--338--1) and Ministry of Education and Science of the Russian Federation (Grant No. 3.1463.2017/4.6).

%
%
%
%

\end{document}